\documentclass[12pt]{article}
\pdfoutput =1
\usepackage{float} 
\usepackage{amsmath}
\usepackage{slashed}
\usepackage{amsfonts}   
\usepackage{amssymb}

\begin{document}
\date{}
\title{{\bf{\Large Analytic integrability for strings on $ \eta $ and $ \lambda $ deformed backgrounds}}}
\author{
 {\bf {\normalsize Dibakar Roychowdhury}$
$\thanks{E-mail:  dibakarphys@gmail.com, Dibakar.RoyChowdhury@swansea.ac.uk}}\\
 {\normalsize  Department of Physics, Swansea University,}\\
  {\normalsize Singleton Park, Swansea SA2 8PP, United Kingdom}
\\[0.3cm]
}

\maketitle
\begin{abstract}
In this paper, based on simple analytic techniques, we explore the integrability conditions for classical stringy configurations defined over $ \eta $ as well as $ \lambda $- deformed backgrounds. We perform our analysis considering classical string motions within various subsectors of the full target space geometry. It turns out that classical string configurations defined over $ \eta $- deformed backgrounds are non-integrable whereas on the other hand, the corresponding configurations are integrable over the $ \lambda $- deformed background. Our analysis therefore imposes a strong constraint on the operator spectrum associated with the corresponding dual gauge theories at strong coupling.
\end{abstract}
\section{Overview and Motivation}
Solving string theory on generic curved backgrounds is indeed a difficult task in itself. Under such circumstances, integrability allows us to choose certain sub-sectors of the theory for which one might at least hope to find out some solution in the same spirit as that with the usual flat space. 
Except for some special circumstances \cite{Alday:2005gi}-\cite{Frolov:2005dj}, it is in general hard to determine whether the given string theory is integrable on generic curved backgrounds under consideration. The key reason for this rests over the fact that there is no generic method and/or prescription for determining Lax pairs on generic curved spaces. However, during the last few years there have been some progress along this particular direction based on both numeric as well as analytic techniques \cite{Frolov:1999pj}-\cite{Basu:2016zkr}.

The key idea associated with the above analytic path rests over the usual notion of integrability in the context of classical Hamiltonian dynamics \cite{Basu:2011fw}-\cite{Stepanchuk:2012xi}. The steps are in general quite straightforward namely, (1) choose an invariant plane in the phase space, (2) obtain the so called normal variational equation (NVE) corresponding to that invariant plane and (3) check whether the NVE admits simple analytic solutions (namely the Liouvillian solutions\footnote{The Liouvillian solutions are essentially simple analytic functions of exponentials, logarithms, simple algebraic expressions involving their integrals \cite{K1}-\cite{K2}.}) using Kovacic's algorithm \cite{K1}-\cite{K2}. The Hamiltonian system is said to be \textit{integrable } in case the corresponding NVE admits Liouvillian solutions.

For the familiarity of the reader, in the following we outline Kovacic's method in brief. Consider a second order differential equation of the following form,
\begin{eqnarray}
a(x) y''(x)+b(x)y'(x)+c(x)y(x)=0\label{D1}
\end{eqnarray} 
where, $ a $, $ b $, $ c $ are all rational functions namely, $ a,b,c \in \mathbb{C}(x) $ where, $ \mathbb{C}(x) $ is the space of rational functions which could be complex in general. It turns out that the above equation (\ref{D1}) is equivalent to the differential equation of the following form \cite{K1}-\cite{K2},
\begin{eqnarray}
\xi ''(x)=\mathfrak{r}(x)\xi (x) \label{D2}
\end{eqnarray}
with, $ \mathfrak{r}(x)=\frac{(2b'a-2ba'+b^{2}-4ac)}{4a^{2}} $. 

Following Kovacic's algorithm \cite{K1}-\cite{K2}, the original differential equation (\ref{D1}) possess a Liouvillian solution if and only if the the solution corresponding to the second equation (\ref{D2}) could be expressed as \cite{K1}-\cite{K2},
\begin{eqnarray}
\xi =e^{\int \omega dx}\label{D3}
\end{eqnarray}
where the function $ \omega (x) $ is algebraic of degree 1, 2, 4, 6 or 12 over $ \mathbb{C}(x) $ \cite{K1}-\cite{K2}. The other way to put this algorithm is to say that the solution (\ref{D3}) exists if and only if the function $ \omega (x) $ satisfying the Ricatti equation,
\begin{eqnarray}
\omega' (x)+\omega^{2} (x)=\mathfrak{r}(x)
\end{eqnarray}
possesses a solution that is algebraic of degree 1, 2, 4, 6 or 12 over $ \mathbb{C}(x) $ \cite{K1}-\cite{K2}.

The purpose of the present article is to apply Kovacic's algorithm to various classical stringy configurations defined over the recently discovered $ \eta $- deformed \cite{Delduc:2013qra}-\cite{Borsato:2016ose} as well as $ \lambda $- deformed backgrounds \cite{Hollowood:2014qma}-\cite{Borsato:2016zcf} and to check the integrability conditions associated with the dynamics of classical string motions over these special classes of target spacetimes those are supposed to preserve integrability in their ground (vacuum) state. As classical strings (are dual to single trace operators in the dual gauge theory at strong coupling) are considered to be excitations above these vacua, therefore the present analysis would give us some indications whether the corresponding operator spectrum (above its ground state) in the dual gauge theory is integrable or not. 

The organization for the rest of the paper is the following: In Section 2, we discuss various stringy configurations both for the $ \eta $ as well as the $ \lambda $ model. In particular, we consider some specific sub-sectors of the full target spacetime and apply Kovacic's algorithm in order to solve stringy dynamics associated with those subspaces. Finally, this article is concluded in Section 3 with some further remarks.
\section{Results}
\subsection{Bosonic strings in $ (AdS_{3}\times S^{3})_{\eta} $}
We begin our discussion with a formal introduction to the bosonic $ \eta $- deformed $ AdS_3 \times S^{3} $ background that could be obtained as a consistent $ 6D $ reduction of the full $ 10D $ solution with the vanishing $ B $ field \cite{Hoare:2014pna},
\begin{eqnarray}
ds^{2}_{AdS_3 \times S^{3}}&=&ds^{2}_{AdS_3}\bigoplus ds^{2}_{S^{3}}\nonumber\\
&=&\left[ -\mathfrak{h}(\varrho)dt^{2}+\mathfrak{f}(\varrho)d\varrho^{2}+\varrho^{2}d\psi^{2}\right] \bigoplus\left[ \tilde{\mathfrak{h}}(\theta)d\varphi^{2}+\tilde{\mathfrak{f}}(\theta)d\theta^{2}+\cos^{2}\theta d\phi^{2}\right] 
\label{E1}
\end{eqnarray}
together with the individual metric functions\footnote{Notice that, here we have introduced a new deformation parameter $ \kappa $ which is related to the original deformation parameter $ \eta $ as, $ \kappa = \frac{2\eta}{1-\eta^{2}}$ \cite{Arutyunov:2013ega}. Therefore, in the subsequent analysis we would always refer the deformation parameter as being $ \kappa $. },
\begin{eqnarray}
\mathfrak{h}&=&\frac{1+\varrho^{2}}{(1-\kappa^{2}\varrho^{2})},~~\mathfrak{f}=\frac{1}{(1+\varrho^{2})(1-\kappa^{2}\varrho^{2})}\nonumber\\
\tilde{\mathfrak{h}}&=&\frac{\sin^{2}\theta}{(1+\kappa^{2}\cos^{2}\theta)},~~\tilde{\mathfrak{f}}=\frac{1}{(1+\kappa^{2}\cos^{2}\theta)}.\label{E2}
\end{eqnarray}

\subsubsection{Strings in $ (R \times S^{3})_{\eta} $}
In this Section, we focus on specific stringy configurations such that strings are sitting at the centre of the $ AdS_3 ~(\varrho =0)$ with non trivial dynamics
associated with the deformed $ S^{3} $. We choose the following ansatz corresponding to the classical string configuration that we are interested in,
\begin{eqnarray}
t=t(\tau),~\theta =\theta (\tau), \varphi (\sigma ,\tau)=\alpha_{1}\sigma +\mathfrak{q}(\tau) , ~\phi = \alpha_{2}\sigma\label{E3}
\end{eqnarray}
where, $ (\tau ,\sigma) $ are the so called world-sheet coordinates. Here $ \alpha_{i} $s are constant coefficients that characterise the winding of the string along angular directions.

The corresponding Polyakov Lagrangian could be formally expressed as\footnote{We have set, $ 2\pi \alpha' =1 $.},
\begin{eqnarray}
\mathcal{L}_{P}=\dot{t}^{2}-\frac{\dot{\theta}^{2}}{1+\kappa^{2}\cos^{2}\theta}+\frac{(\alpha_{1}^{2}-\dot{\mathfrak{q}}^{2})\sin^{2}\theta}{1+\kappa^{2}\cos^{2}\theta}+\alpha_{2}^{2}\cos^{2}\theta. \label{E4}
\end{eqnarray}

The non trivial set of equations that readily follow from (\ref{E4}) could be expressed as,
\begin{eqnarray}
\ddot{\theta}(1+\kappa^{2}\cos^{2}\theta)+\kappa^{2}\dot{\theta}^{2}\sin\theta \cos\theta +(\alpha_{1}^{2}-\dot{\mathfrak{q}}^{2})(\sin\theta \cos\theta (1+\kappa^{2}\cos^{2}\theta)+\kappa^{2}\sin^{3}\theta \cos\theta)\nonumber\\
-\alpha^{2}_{2}\sin\theta \cos\theta (1+\kappa^{2}\cos^{2}\theta)^{2}=0\nonumber\\
\ddot{\mathfrak{q}}\sin^{2}\theta (1+\kappa^{2}\cos^{2}\theta)+2\dot{\mathfrak{q}}\dot{\theta}(\sin\theta \cos\theta (1+\kappa^{2}\cos^{2}\theta)+\kappa^{2}\sin^{3}\theta \cos\theta)=0.\label{E5}
\end{eqnarray}

With the above equation (\ref{E5}) in hand, we consider $ \theta,\dot{\theta}\rightarrow 0 $ limit of the second equation in (\ref{E5}). Inorder to do that, we first set,
\begin{eqnarray}
\theta \sim \dot{\theta}\sim \epsilon \label{E6}
\end{eqnarray}
such that $ |\epsilon|\ll 1 $. This eventually defines our invariant plane $ \lbrace \theta \sim 0 , p_{\theta}\sim 0\rbrace $ in the phase space \cite{Basu:2011fw}-\cite{Stepanchuk:2012xi}.

Substituting (\ref{E6}) into (\ref{E5}) and taking the limit, $ \epsilon \rightarrow 0 $ we finally obtain,
\begin{eqnarray}
\ddot{\mathfrak{q}}+2\dot{\mathfrak{q}} \approx 0
\end{eqnarray}
which possesses a classical solution of the form, $ \mathfrak{q}=\mathfrak{q}_{s} (\tau)=e^{-2\tau}$. 

Our next task would be to explore whether small perturbations around this special solution is integrable in the sense of Kovacic's algoritm \cite{K1}-\cite{K2} of finding Liouvillian solution.  The corresponding NVE is obtained by considering small fluctuations, $ \theta \sim \eta (\tau),~|\eta|\ll 1 $ around the above solution which for the present case yields,
\begin{eqnarray}
\ddot{\eta}+(\alpha_{1}^{2}-\dot{\mathfrak{q}}_{s}^{2}-\alpha^{2}_{2}(1+\kappa^{2}))\eta =0.\label{E8}
\end{eqnarray}

In order to check Kovacic's algorithm, we consider the following change of variables namely,
\begin{eqnarray}
z=e^{-4\tau}.\label{E9}
\end{eqnarray}

Using (\ref{E9}), we finally obtain,
\begin{eqnarray}
z^{2}\eta''(z)+z\eta'(z)+\mathfrak{K}(z)\eta (z)=0\label{E10}
\end{eqnarray}
where, $ \mathfrak{K}(z)=\frac{(\alpha^{2}_{1}-4z-\alpha^{2}_{2}(1+\kappa^{2}))}{16} $.

Clearly, the above equation (\ref{E10}) is a simple homogeneous linear second order differential equation with Polynomial coefficients which thereby allows us to check Kovacic's algorithm \cite{K1}-\cite{K2} directly. For generic values of the parameters, the solution could be formally expressed as,
\begin{eqnarray}
\eta (z)=\mathcal{C}_{1}(-1)^{-\frac{1}{4} \sqrt{\alpha_{2}^2 \left(\kappa ^2+1\right)-\alpha_{1}^2}}  \Gamma \left(1-\frac{1}{2} \sqrt{\alpha_{2}^2 \left(\kappa ^2+1\right)-\alpha_{1}^2}\right) I_{-\frac{1}{2} \sqrt{\alpha_{2}^2 \left(\kappa ^2+1\right)-\alpha_{1}^2}}\left(\sqrt{z}\right)\nonumber\\+\mathcal{C}_{2}(-1)^{-\frac{1}{4} \sqrt{\alpha_{2}^2 \left(\kappa ^2+1\right)-\alpha_{1}^2}} i^{\sqrt{\alpha_{2}^2 \left(\kappa ^2+1\right)-\alpha_{1}^2}} \Gamma \left(\frac{1}{2} \sqrt{{\alpha_{2}}^2 \left(\kappa ^2+1\right)-\alpha_{1}^2}+1\right) I_{\frac{1}{2} \sqrt{\alpha_{2}^2 \left(\kappa ^2+1\right)-\alpha_{1}^2}}\left(\sqrt{z}\right)
\end{eqnarray}
where, $ I_{a}(x) $ is the modified Bessel's function of the first kind and $ \Gamma (n) $ is the usual Gamma function.

Following the Kovacic's algorithm\cite{K1}-\cite{K2}, an equivalent equation corresponding to (\ref{E8}) could be formally expressed as,
\begin{eqnarray}
\xi ''(z)=\left(\frac{-\alpha_{1}^{2}+\alpha_{2}^{2} \left(\kappa ^2+1\right)+4 z-4}{16 z^2} \right) \xi (z)=\mathfrak{r}(z)\xi (z).\label{E12}
\end{eqnarray}

The solution corresponding to (\ref{E12}) turns out to be,
\begin{eqnarray}
\xi (z)=\frac{\mathcal{C}_{1}}{2} i \sqrt{z} (-1)^{-\frac{1}{4} \sqrt{\alpha_{2}^2 \left(\kappa ^2+1\right)-\alpha_{1}^2}}  \Gamma \left(1-\frac{1}{2} \sqrt{\alpha_{2}^2 \left(\kappa ^2+1\right)-\alpha_{1}^2}\right) I_{-\frac{1}{2} \sqrt{\alpha_{2}^2 \left(\kappa ^2+1\right)-\alpha_{1}^2}}\left(\sqrt{z}\right)\nonumber\\
+\frac{\mathcal{C}_{2}}{2} i \sqrt{z} (-1)^{-\frac{1}{4} \sqrt{\alpha_{2}^2 \left(\kappa ^2+1\right)-\alpha_{1}^2}} i^{\sqrt{\alpha_{2}^2 \left(\kappa ^2+1\right)-\alpha_{1}^2}} \Gamma \left(\frac{1}{2} \sqrt{\alpha_{2}^2 \left(\kappa ^2+1\right)-\alpha_{1}^2}+1\right) I_{\frac{1}{2} \sqrt{\alpha_{2}^2 \left(\kappa ^2+1\right)-\alpha_{1}^2}}\left(\sqrt{z}\right)
\end{eqnarray}

If we now demand that the above equation (\ref{E12}) has a solution which is of the form,
\begin{eqnarray}
\xi (z)=e^{\int \omega (z)dz}
\end{eqnarray}
then the corresponding function $ \omega (z) $ turns out to be,
\begin{eqnarray}
\omega (z)=\frac{\mathcal{N}(z)}{4z\mathcal{D}(z)}\label{E15}
\end{eqnarray}
where each of the individual functions could be formally expressed as,
\begin{eqnarray}
\mathcal{N}(z)= \Gamma \left(1-\frac{\mathfrak{n}}{2} \right) 2 \sqrt{z} I_{1-\frac{\mathfrak{n}}{2} }\left(\sqrt{z}\right)-\left(\mathfrak{n}-2\right) I_{-\frac{\mathfrak{n}}{2} }\left(\sqrt{z}\right)\nonumber\\+2 e^{\frac{\mathfrak{i \pi n}}{2}  }\left( \sqrt{z} I_{\frac{1}{2} \left(\mathfrak{n}+2\right)}\left(\sqrt{z}\right)\Gamma \left(\frac{\mathfrak{n}}{2} +1\right)+\Gamma \left(\frac{\mathfrak{n}}{2} +2\right) I_{\frac{\mathfrak{n}}{2} }\left(\sqrt{z}\right)\right) 
\end{eqnarray}
\begin{eqnarray}
\mathcal{D}(z)=  \Gamma \left(1-\frac{\mathfrak{n}}{2} \right) I_{-\frac{\mathfrak{n}}{2} }\left(\sqrt{z}\right)+i^{\mathfrak{n}} \Gamma \left(\frac{\mathfrak{n}}{2} +1\right) I_{\frac{\mathfrak{n}}{2} }\left(\sqrt{z}\right)
\end{eqnarray}
\begin{eqnarray}
\mathfrak{n}=\sqrt{\alpha_{2}^2 \left(\kappa ^2+1\right)-\alpha_{1}^2}.
\end{eqnarray}

It turns out that for generic $ \mathfrak{n} $ and $ |z|<1 $ (\ref{E9}), the function $ \omega (z) $ is indeed a Polynomial of the following form,
\begin{eqnarray}
\omega (z)=\frac{\mathcal{C}_{1}(\mathfrak{n})}{z}+\frac{\mathcal{C}_{2}(\mathfrak{n})}{z^{1-\frac{\mathfrak{n}}{2}}}+\frac{\mathcal{C}_{3}(\mathfrak{n})}{z^{1-\mathfrak{n}}}+..~..
\end{eqnarray}
where, $ \mathcal{C}_{i} $s are some complex coefficients in general.

As a special case, for $ \mathfrak{n}=0 $ (which in turn fixes, $ \kappa^{2}=\frac{\alpha_{1}^{2}-\alpha_{2}^{2}}{\alpha_{2}^{2}} $) we find,
\begin{eqnarray}
\omega (z)=\frac{1}{2 z}+\frac{1}{4}-\frac{z}{32}+\frac{z^2}{192}+..\label{E19}
\end{eqnarray}

Finally, after a straightforward computation, it turns out that the corresponding Ricatti equation \cite{K1}-\cite{K2} is also satisfied,
\begin{eqnarray}
\omega' (z)+\omega^{2}(z)=\mathfrak{r}(z).\label{E20}
\end{eqnarray}

For the special case with, $ \mathfrak{n}=0 $, Eq.(\ref{E19}) is the corresponding (Polynomial) solution to the Ricatti  equation (\ref{E20}). 
From the above analysis it is therefore quite evident that the original NVE (\ref{E8}) we started with does not possess Liouvillian solution and therefore the corresponding stringy configuration (\ref{E3}) (and hence the dual field theory) is non-integrable.
\subsubsection{Pulsating strings in $ (AdS_{3})_{\eta} $}
The purpose of this Section is to check integrability conditions for classical pulsating strings over $ \kappa $- deformed $ AdS_3 $ \cite{Panigrahi:2014sia},
\begin{eqnarray}
ds^{2}=-\frac{\cosh^{2}\varrho}{1-\kappa^{2}\sinh^{2}\varrho}dt^{2}+\frac{d\varrho^{2}}{1-\kappa^{2}\sinh^{2}\varrho}+\sinh^{2}\varrho d\psi^{2}.\label{E22}
\end{eqnarray}
We now wish to check the integrability corresponding to the ansatz of the following form,
\begin{eqnarray}
t=t(\tau),~\varrho=\varrho(\tau), ~\psi= \alpha \sigma.\label{E23}
\end{eqnarray}

Next, we note down the corresponding non trivial equations of motion that readily follow from (\ref{E23}),
\begin{eqnarray}
\ddot{t}\cosh^{2}\varrho +\dot{\varrho}\dot{t}\sinh 2\varrho \left( 1+\frac{\kappa^{2}\cosh^{2}\varrho}{1-\kappa^{2}\sinh^{2}\varrho}\right) &=&0\nonumber\\
2\ddot{\varrho}(1-\kappa^{2}\sinh^{2}\varrho)+\sinh\varrho (\alpha^{2}(1-\kappa^{2}\sinh^{2}\varrho)^{2}+\kappa^{2}\dot{\varrho}^{2}+\dot{t}^{2}(1+\kappa^{2}))&=&0.
\end{eqnarray}

The invariant plane in the phase space is set by the following ansatz,
\begin{eqnarray}
\varrho=\dot{\varrho}=0 \label{E25}
\end{eqnarray}
which yields,
\begin{eqnarray}
t=\gamma \tau.
\end{eqnarray}

As usual, our next task would be to consider fluctuations ($ \delta\varrho \sim |\vartheta (\tau)|\ll 1 $) normal to the plane (\ref{E25}) explore NVE in the sense of Kovacic's algorithm \cite{K1}-\cite{K2}. The NVE for the present example turns out to be,
\begin{eqnarray}
2\ddot{\vartheta}+(\alpha^{2}+\gamma^{2}(1+\kappa^{2}))\vartheta =0
\end{eqnarray}
which is nothing but the equation corresponding to a simple harmonic oscillator and hence the original stringy configuration is trivially integrable \cite{Basu:2011fw}. This seems to be the only configuration that might preserve integrability.

\subsubsection{Spiky strings in $ (AdS_{3})_{\eta} $}
The purpose of this Section is to study the integrability conditions associated with the spiky string configurations \cite{Kruczenski:2004wg}-\cite{Banerjee:2015nha} over $ \kappa $- deformed $ AdS_3 $ (\ref{E2}). The configuration that we are interested in is the following,
\begin{eqnarray}
t=\tau,~\varrho =\varrho (\tau),~\psi (\sigma , \tau)=\beta \sigma +\mathfrak{s}(\tau).\label{E28}
\end{eqnarray}

The corresponding Polyakov Lagrangian could be formally expressed as,
\begin{eqnarray}
\mathcal{L}_{P}=\frac{(1+\varrho^{2})}{(1-\kappa^{2}\varrho^{2})}-\frac{\dot{\varrho}^{2}}{(1+\varrho^{2})(1-\kappa^{2}\varrho^{2})}+\varrho^{2}(\beta^{2}-\dot{\mathfrak{s}}^{2})
\end{eqnarray}
which yields the following set of non trivial equations,
\begin{eqnarray}
\ddot{\varrho}(1+\varrho^{2})(1-\kappa^{2}\varrho^{2})-\dot{\varrho}^{2}\varrho (1-\kappa^{2}-2\kappa^{2}\varrho^{2})+\varrho (1+\kappa^{2})(1+\varrho^{2})^{2}\nonumber\\ +2\varrho (1+\varrho^{2})^{2}(1-\kappa^{2}\varrho^{2})^{2}(\beta^{2}-\dot{\mathfrak{s}}^{2})&=&0\nonumber\\
\varrho \ddot{\mathfrak{s}}+2\dot{\varrho}\dot{\mathfrak{s}}&=&0.
\end{eqnarray}

The invariant plane in the phase space is fixed by the following ansatz,
\begin{eqnarray}
\lbrace\varrho , \dot{\varrho} \rbrace\approx 0
\end{eqnarray}
which yields,
\begin{eqnarray}
\mathfrak{s}(\tau)=\mathfrak{s}_{c}=e^{-2\tau}.
\end{eqnarray}

The corresponding NVE takes the following form,
\begin{eqnarray}
\ddot{\vartheta}+(1+\kappa^{2}+2(\beta^{2}-\dot{\mathfrak{s}}_{c}^{2}))\vartheta =0.
\end{eqnarray}

As a next step of our analysis, we set,
\begin{eqnarray}
z=e^{-4\tau}
\end{eqnarray}
which yields,
\begin{eqnarray}
z^{2}\vartheta''(z)+z\vartheta'(z)+\mathfrak{G}(z)\vartheta (z)=0\label{E35}
\end{eqnarray}
where, $ \mathfrak{G}(z)=\frac{(1+\kappa^{2}+2\beta^{2}-8z)}{16} $. Clearly, the above equation (\ref{E35}) possesses remarkable similarity to that with the previously obtained equation (\ref{E10}) which thereby sort of guarantee the non-integrability corresponding to the above stringy configuration (\ref{E28}).

\subsection{Bosonic strings in $  (AdS_{3}\times S^{3})_{\lambda}$}
The target space metric corresponding to $ \lambda $ deformations could be formally expressed as \cite{Hoare:2015gda},
\begin{eqnarray}
2\pi k^{-1}ds^{2}=\frac{1}{1+2b^{2}}\left( -dt^{2}+J^{2}+\coth^{2}\xi K^{2}-4b^{2}(1+b^{2})(\cosh^{2}\xi(dt-K)^{2}-J^{2})\right) \nonumber\\
+\frac{1}{1+2b^{2}}\left(d\varphi^{2}+\tilde{J}^{2}+\cot^{2}\zeta \tilde{K}^{2}+4b^{2}(1+b^{2})(\cos^{2}\zeta(d\varphi +\tilde{K})^{2}+\tilde{J}^{2}) \right) 
\end{eqnarray}
where the individual metric functions could be formally expressed as, 
\begin{eqnarray}
J&=&\csc(2t)(\sin(2\psi)d\xi -\coth\xi(\cos(2t)-\cos(2\psi))d\psi)\nonumber\\
K&=&\csc(2t)(\tanh\xi (\cos(2t)+\cos(2\psi))d\xi -\sin(2\psi)d\psi)\nonumber\\
\tilde{J}&=&\csc(2\varphi)(\sin(2\phi)d\zeta +\cot\zeta(\cos(2\varphi)-\cos(2\phi))d\phi)\nonumber\\
\tilde{K}&=&\csc(2\varphi)(\tan\zeta (\cos(2\varphi)+\cos(2\phi))d\zeta +\sin(2\phi)d\phi).
\end{eqnarray}
Here, the parameter $ b $ is related to the original deformation parameter $ \lambda $ as \cite{Hoare:2015gda},
\begin{eqnarray}
b^{2}=\frac{\lambda^{2}}{1-\lambda^{2}}, ~\lambda^{2}\in [0,1].
\end{eqnarray}

\subsubsection{Strings in $ (R \times S^{2})_{\lambda} $}
To check integrability, we would consider a simple (sub)sector of the theory namely, we would be considering strings moving in  $ (R \times S^{2})_{\lambda} $.
The corresponding stringy ansatz in the bulk turns out to be,
\begin{eqnarray}
t=\tau, ~\xi =\xi_{c}, ~\psi = \psi_{c},~\zeta = \zeta_{c}=\frac{\pi}{4},~\varphi =\varphi(x),~\phi = \phi (x)\label{E39}
\end{eqnarray}
where, the coordinate $ x $ stands for either of the worldsheet coordinates ($ \tau ,\sigma $).

With the above choice (\ref{E39}), the corresponding metric simplifies to,
\begin{eqnarray}
2\pi k^{-1}ds^{2}=g_{tt}dt^{2}+g_{\varphi\varphi}d\varphi^{2}+g_{\phi \phi} d\phi^{2} +2g_{\phi \varphi}d\phi d\varphi
\end{eqnarray}
where the individual metric functions could be formally expressed as\footnote{We ignore the overall scaling factor, $ \frac{1}{1+2b^{2}} $.},
\begin{eqnarray}
g_{tt}&=&-(1+4b^{2}(1+b^{2})\cosh^{2}\xi_{c})\nonumber\\
g_{\varphi\varphi}&=&(1+2b^{2}(1+b^{2}))\nonumber\\
g_{\phi\varphi}&=&2b^{2}(1+b^{2})\csc (2\varphi)\sin (2\phi)
\end{eqnarray}
and,
\begin{eqnarray}
g_{\phi \phi}=\csc^{2}(2\varphi)(1+2b^{2}(1+b^{2}))(\cos^{2}2\varphi -2\cos 2\varphi \cos 2\phi +1)  \nonumber\\
+2b^{2}(1+b^{2})\csc^{2}(2\varphi)(\cos 2\varphi -\cos 2\phi)^{2}.
\end{eqnarray}

The corresponding Polyakov Lagrangian could be formally expressed as\footnote{Here, the dot corresponds to derivative w.r.t. the worldsheet coordinate $ x $ which could be $ \tau $ or $ \sigma $.},
\begin{eqnarray}
\mathcal{L}_{P}=g_{\varphi\varphi}\dot{\varphi}^{2}+g_{\phi\phi}\dot{\phi}^{2}+g_{\varphi \phi}\dot{\varphi}\dot{\phi}
\end{eqnarray}
which yields the following set of non trivial equations,\\
$ \bullet $ \underline{\textbf{$ \varphi  $ Equation:}} 
\begin{eqnarray}
(1+2b^{2}(1+b^{2}))\ddot{\varphi}+2\dot{\phi}^{2}\csc^{2}(2\varphi)(\left(b^4+b^2\right) \cos (4 \phi)+7 \left(b^4+b^2\right)+2)\nonumber\\
+\dot{\phi}^{2}\csc^{2}(2\varphi)\left(\cot (2\varphi)-\left(2 b^2+1\right)^2 (\cos (4 \varphi)+3) \csc (2 \varphi) \cos (2 \phi) \right) 
+b^{2}(1+b^{2})\ddot{\phi}\csc(2\varphi)\sin(2\phi)=0
\end{eqnarray}
$ \bullet $ \underline{\textbf{$ \phi $ Equation:}} 
\begin{eqnarray}
g_{\phi\phi}\ddot{\phi}+2\csc (2 \varphi) \sin (2 \phi) \left(\left(2 b^2+1\right)^2 \cot (2 \varphi)-2 b^2 \left(b^2+1\right) \csc (2 \varphi) \cos (2 \phi)\right)\dot{\phi}^{2}\nonumber\\
+b^{2}(1+b^{2})\ddot{\varphi}\csc (2\varphi)\sin (2\phi)=0.
\end{eqnarray}

We set the invariant plane in the phase space with the ansatz,
\begin{eqnarray}
\phi =0,~\dot{\phi}=0\label{E46}
\end{eqnarray}
which yields,
\begin{eqnarray}
\varphi (x)=\varphi_{c}=\mathfrak{c}x.
\end{eqnarray}

The corresponding NVE turns out to be,
\begin{eqnarray}
\ddot{\eta}\approx 0\label{E48}
\end{eqnarray}
where, as usual $ \eta (x) $ is the corresponding fluctuation in the direction normal to the invariant plane (\ref{E46}) in the phase space. Eq.(\ref{E48}) is remarkably simple and admits simple analytic (Liouvillian) solution,
\begin{eqnarray}
\eta(x)=x
\end{eqnarray}
which thereby (unlike the case for the $ \eta $- deformations) preserves the integrability of the corresponding stringy configuration (\ref{E39}).
\subsubsection{Strings in $ (AdS_{2})_{\lambda} $}
We now aim to explore integrability for strings moving in another subsector of the full theory namely the $ \lambda $- deformed $ AdS_2 $. We choose the following ansatz for the stringy configuration,
\begin{eqnarray}
t=t(\tau),~\psi =\psi (\tau),~\xi =\xi_{c}=\coth^{-1}1.\label{E50}
\end{eqnarray}

With the above ansatz (\ref{E50}), the corresponding metric (ignoring all the overall scale factors) simplifies to,
\begin{eqnarray}
ds^{2}=-dt^{2}+g_{\psi\psi}d\psi^{2}
\end{eqnarray}
where, the corresponding metric function could be expressed as,
\begin{eqnarray}
g_{\psi\psi}=\csc^{2}(2t)(\cos^{2}(2t)-2\cos(2t)\cos(2\psi)+1)+4b^{2}(1+b^{2})\csc^{2}(2t)(\cos(2t)-\cos(2\psi))^{2}.
\end{eqnarray}

The corresponding Polyakov action turns out to be,
\begin{eqnarray}
\mathcal{L}_{P}=\dot{t}^{2}-g_{\psi\psi}\dot{\psi}^{2}
\end{eqnarray}
which yields the following set of equations of motion,
\begin{eqnarray}
\ddot{t}+\csc ^2(2 t) \left(\left(2 b^2+1\right)^2 (\cos (4 t)+3) \csc (2 t) \cos (2 \psi)\right)\dot{\psi}^{2}\nonumber\\
-4 \csc ^2(2 t)\cot (2 t) \left(\left(b^4+b^2\right) \cos (4 \psi)+3 \left(b^4+b^2\right)+1\right)\dot{\psi}^{2}=0
\end{eqnarray}
and,
\begin{eqnarray}
g_{\psi\psi}\ddot{\psi}+2 \csc (2 t) \sin (2 \psi) \left(4 \left(b^2+1\right) b^2 \csc (2 t) (\cos (2 t)-\cos (2 \psi))+\cot (2 t)\right)\dot{\psi}^{2}=0.
\end{eqnarray}

In the following we look for an invariant plane in the phase space which is obtained by setting up an ansatz of the following form namely,
\begin{eqnarray}
\psi =\dot{\psi}=0
\end{eqnarray}
which yields,
\begin{eqnarray}
t(\tau)=\tau.
\end{eqnarray}

Finally, the corresponding NVE associated with the fluctuations, $ \delta\psi \sim \vartheta (\tau)$ tuns out to be extremely simple ,
\begin{eqnarray}
\ddot{\vartheta}\approx 0
\end{eqnarray}
which clearly admits Liouvillian solution and hence the corresponding stringy configuration (\ref{E50}) is trivially integrable.

\section{Summary and final remarks}
We now summarize our analysis. The purpose of the present analysis was to explore the integrability conditions corresponding to classical stringy configurations defined over the newly discovered $ \eta $ as well as $ \lambda $- deformed backgrounds. In our analysis, we stick to the bosonic sector of the full super-string target space and rather than considering the dynamics over the full background geometry, we consider strings moving in various sub-sectors of the full target space. 

Our analysis reveals that the stringy motion over $ \eta $- deformed backgrounds are non-integrable. On the other hand, the classical string configurations defined over $ \lambda $- deformed backgrounds turn out to be integrable. Since classical strings correspond to single trace operators in the dual gauge theory, therefore our analysis imposes severe constraints on the corresponding operator spectrum at strong coupling. Our analysis corresponds to the fact that the excitations (above the ground state) associated with the gauge theory dual to $ \eta $- deformations do not preserve any integrable structure, whereas on the other hand, the corresponding operator spectrum turns out to be integrable for $ \lambda $- deformations.\\ \\
{\bf {Acknowledgements :}}
This work was supported through the Newton-Bhahba Fund. The author would like to acknowledge the Royal Society UK and the Science and Engineering Research Board India (SERB) for financial assistance. The author would also like to acknowledge Prof. Carlos Nunez for helpful discussions. \\ 

\end{document}